# LaAlO$_3$/SrTiO$_3$-Based Device Concepts

Daniela F. Bogorin, Patrick Irvin, Cheng Cen, Jeremy Levy

*Department Of Physics and Astronomy, University Of Pittsburgh*

*Pittsburgh, PA 15260*

## Introduction

The two-dimensional electron gas (2DEG) that forms at the interface between two heterogeneous semiconductors or between a semiconductor and oxide is the basis for some of the most useful and prevalent electronic devices. Silicon-based metal-oxide-semiconductor field-effect transistors, GaAs-based high-mobility transistors, solid-state lasers and photodetectors are but a few examples of technologies that have emerged from semiconductor interfaces. Herb Kroemer, when receiving the Nobel Prize in Physics for the invention of the semiconductor heterostructure, stated that "the interface is the device" [1].

Complex oxides differ in many ways from traditional semiconductors that are used to form 2DEG devices [2]. Both LaAlO$_3$ and SrTiO$_3$ are compatible with a large class of materials that, collectively, exhibits a remarkable range of behaviors that include ferroelectricity, magnetism, multiferroic behavior, and superconductivity. There are many other complex oxide systems with exceptional properties, specifically ZnO/MgZnO heterostructures grown with high mobility [3].

The development of complex oxides over the past fifteen years has raised the prospect for new classes of electronic devices [2, 4]. In 2004, Ohtomo and Hwang published their seminal discovery that a high-mobility 2DEG can form at the interface between LaAlO$_3$ and SrTiO$_3$ [5], which forms the backdrop for this chapter. Since then a number of striking properties of this interface have been discovered and explored. Here we will highlight the most important features that may lead to novel devices that can challenge or disrupt current technologies and perhaps create new ones. One feature in particular that we will focus on is the ability to reversibly create conducting nanostructures at the LaAlO$_3$/SrTiO$_3$ interface using a conductive atomic-force microscope (c-AFM). Again, it seems appropriate to quote from Kroemer's Nobel Lecture, in which he introduced his "*Lemma of New Technology*" [1]:

> *The principal applications of any sufficiently new and innovative technology always have been – and will continue to be – applications created by that technology.*

In the chapter below we will describe features and functionality that *resemble* existing technologies, but we hope that the reader will be inspired to think about what kind of *fundamentally new* technologies may also emerge.

## Semiconductor 2DEGs

A 2DEG is a structure in which the electrons are restricted to move only in a two-dimensional plane. Quantum confinement along the growth direction leads to the formation of "subbands," in which the lowest one is typically occupied for a 2DEG. Semiconductor 2DEGs are traditionally formed in Si metal-oxide-semiconductor field-effect transistors (MOSFETs) or modulation-doped GaAs/AlGaAs heterostructures. In modulation-doped structures the carriers are physically separated from the doping centers, enabling very high mobilities up to 31,000,000 $cm^2$/Vs [6].

Graphene is another 2D system of contemporary interest [7]. A fascinating property of the Dirac fermions in graphene is their ability to "Klein tunnel" through large barriers without scattering [8]. This property leads to anomalously large mobilities at room temperature.

## 2DEG at $LaAlO_3$/$SrTiO_3$ interface

$LaAlO_3$ and $SrTiO_3$ both have the perovskite crystal structure. $SrTiO_3$, a non-polar oxide, is composed of alternating, stacked layers of $(SrO)^0$ and $(TiO_2)^0$ (Figure 1). $SrTiO_3$ is pseudo-cubic at room temperature with a lattice constant 3.905 Å. It is a band-insulator with a bandgap of 3.25 eV, is chemically inert, and has been a substrate of choice for the growth of many other oxides and high-$T_C$ superconductors [4, 9, 10]. $LaAlO_3$, a polar oxide, is composed of alternating stacked layers of $(LaO)^+$ and $(AlO_2)^-$. $LaAlO_3$ has a pseudo-cubic structure with a lattice constant of 3.791 Å at room temperature. It is a Mott insulator [11] with a wide bandgap of 5.6 eV. Films of $LaAlO_3$ thinner than ~15 monolayers (ML) [12] can be grown coherently strained to $SrTiO_3$. The manageably small lattice mismatch between $LaAlO_3$ and $SrTiO_3$ (about 3%) enables high-quality epitaxial growth of $LaAlO_3$ on $SrTiO_3$ via pulsed laser deposition (PLD) [13] or molecular-beam epitaxy (MBE) [14].

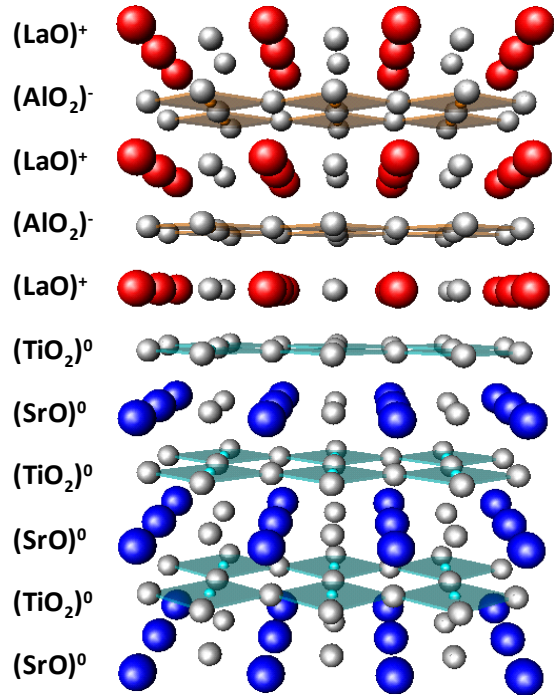

Figure 1. LaAlO$_3$/(TiO$_2^-$)SrTiO$_3$ structure. (Adapted from Ref. [13].)

## Polar catastrophe model

Epitaxial growth of LaAlO$_3$ on SrTiO$_3$ [5] can lead to an unusual and energetically unstable charge distribution. Two situations are shown in Figure 2, the difference being the final termination of the SrTiO$_3$ layer. We first consider TiO$_2$-terminated SrTiO$_3$ (Figure 2 (a)). The first-grown LaO layer will have a positive charge (+1/unit cell), after which the charge of each layer alternates between -1 and +1. This alternating charge density can be integrated to reveal both the internal electric field $E$ and voltage $V$ across the LaAlO$_3$ layer. The magnitude of the voltage increases linearly with the thickness of the LaAlO$_3$ layer. The divergence of the electrostatic energy can lead to a "polar catastrophe" and an associated electronic reconstruction in which the polarization is screened (in part) by the formation of a 2DEG at the LaAlO$_3$/SrTiO$_3$ interface (Figure 2 (c)). The other possible growth condition (Figure 2 (b)) occurs when the SrTiO$_3$ substrate is SrO-terminated. This situation also leads to a "polar catastrophe," although it is predicted to result in a positively-charged interface.

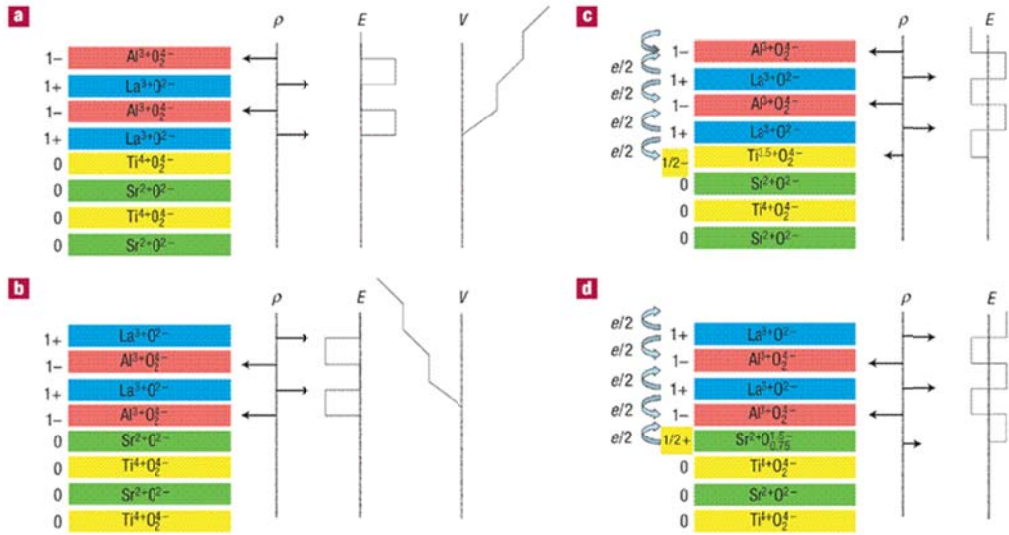

Figure 2. Illustration of charge redistribution due to polar catastrophe at interfaces of $LaAlO_3/SrTiO_3$ grown with different termination conditions. (a) Non-reconstructed $LaO/TiO_2$ interface. A positive potential exists in the $LaAlO_3$ layer and diverges with film thickness. (b) A similar but negative potential occurs in the case of an $AlO_2/SrO$ interface. (c) At the $LaO/TiO_2$ interface, adding 1/2 electron per unit cell to the top $TiO_2$ layer can minimize the potential in $LaAlO_3$. (d) At the $AlO_2/SrO$ interface, adding 1/2 hole per unit cell to the top SrO layer can minimize the potential in $LaAlO_3$. (Adapted from Ref. [5].)

One missing component of this model is a discussion of what happens at the top $LaAlO_3$ surface. The absence of charge neutrality (indicated in Figure 2 (c,d)) leads to a much more severely catastrophic situation than the existence of a finite, but diverging, polarization. To achieve charge neutrality, the top surface must also be compensated with charge. One such scenario that was invoked to explain metastable writing at the $LaAlO_3/SrTiO_3$ interface (described in more detail below) is the formation of oxygen vacancies at the top $LaAlO_3$ surface. Another scenario, described by Son et al [15], involves the adsorption of hydrogen at the top $AlO_2$ surface. The hydrogen could be formed by spontaneous dissociation of $H_2O$ molecules or through direct adsorption of $H_2$ in gas form. The charged top surface is expected to retain its insulating properties, but donation of electrons to the $LaAlO_3/SrTiO_3$ interface can be achieved for $TiO_2$-terminated $SrTiO_3$.

## Metal-insulator transition in $LaAlO_3/SrTiO_3$

One predicted consequence of the polar discontinuity between $LaAlO_3$ and $SrTiO_3$ is an interfacial insulator-to-metal transition that is dependent on the $LaAlO_3$ thickness. This effect, first reported in 2006 by Thiel et al [13] and confirmed by many groups, occurs in structures of $LaAlO_3$ grown on $TiO_2$-terminated $SrTiO_3$. When $LaAlO_3$ is grown with greater than a critical thickness $d_c$ = 3 unit cell (uc), the interface between $LaAlO_3$ and $SrTiO_3$ is found to be conducting. When the thickness of $LaAlO_3$ is smaller than $d_c$ the interface remains insulating. In samples grown with approximately 3 uc of LaAlO$_3$ (normally insulating), the interface can be switched between the insulating and conducting states by applying a voltage to the back of the SrTiO$_3$ substrate [13]. This transition is a hysteretic function of the applied electric field.

**Inconsistencies with polar catastrophe model**

While the polar catastrophe model helps to explain a number of empirical observations in the LaAlO$_3$/SrTiO$_3$ system, several experiments have identified inconsistencies with this model. Typical mobilities of a 2DEG formed at the LaAlO$_3$/SrTiO$_3$ interface is on the order of 100-1000 cm$^2$/Vs. However, in the original paper by Ohtomo and Hwang [5], the carrier density of the highest mobility LaAlO$_3$/SrTiO$_3$ structure ($\mu = 10^4$ cm$^2$/Vs at 2 K) was found to be two orders of magnitude larger than that required to screen the polarization in the LaAlO$_3$. The most likely explanation is that the majority of carriers in this system were donated by other sources (e.g., oxygen vacancies). Subsequent investigations with samples grown under different conditions have yielded carrier densities that are lower, but consistent with the polar catastrophe model.

The effect of oxygen vacancies has been investigated explicitly by several groups. During growth of LaAlO$_3$ by PLD, the growth temperature, oxygen pressure, and post-deposition annealing conditions can greatly influence the distribution of oxygen vacancies [16, 17]. In addition, substrate preparation methods [18] used to produce a TiO$_2$-terminated surface can lead to the formation of oxygen vacancies. Under certain conditions, the density of oxygen vacancies can be sufficient to lead to three-dimensional electronic properties (i.e., no confinement) near the LaAlO$_3$/SrTiO$_3$ interface [19-23]. Post-deposition annealing can significantly reduce the density of oxygen vacancies near the interface, as has been measured directly using cross-sectional conducting AFM [21]. Because oxygen vacancies are charged they are capable of redistribution under applied electric fields. The hysteretic switching of 3uc-LaAlO$_3$/SrTiO$_3$ observed by Thiel *et al* [13] can be explained by the field-induced hysteretic motion of oxygen vacancies at or near the LaAlO$_3$/SrTiO$_3$ interface.

Another experimental observation that is difficult to reconcile with the polar catastrophe model is the absence of clear signature in x-ray photoemission spectroscopy (XPS) measurements [24-27]. A number of alternate theoretical explanations have been given to describe the interface confinement [28], however at this point a complete physical picture has yet to emerge.

# Field-Effect Devices

The variety of electronic phenomena exhibited at the LaAlO$_3$/SrTiO$_3$ interface depends on the ability to modulate the carrier density at the interface through a combination of electronic reconstruction, doping, and field effects. Electronic reconstruction was discussed above and can drive the LaAlO$_3$/SrTiO$_3$ interface close to or through the metal-insulator transition. Direct chemical doping of SrTiO$_3$ with Nb can produce high-mobility 2DEGs that become superconducting at low temperatures [29, 30]. Oxygen vacancies can donate electrons to the interface if they are located at or near the interface itself. The top LaAlO$_3$ surface, or even defects within the LaAlO$_3$, can also provide a kind of modulation doping of the LaAlO$_3$/SrTiO$_3$ interface.

## SrTiO$_3$-based channels

A variety of field-effect devices that use SrTiO$_3$ as the channel layer have been reported [31-36]. The SrTiO$_3$ channel can be either intrinsic or heavily doped. One device utilized the ferroelectric field effect from a Pb (Zr$_{0.2}$Ti$_{0.8}$)O$_3$ (PZT) to modulate the superconducting transition temperature of a 400 Å, Nb-doped SrTiO$_3$ using the metallic tip of an AFM as a mobile gate electrode [34]. Similar approaches were reported in Refs. [31, 33]. FET devices were created by combining two p-type copper oxide superconductors gated by the SrTiO$_3$ dielectric layer. A metal-insulator-semiconductor field effect transistor (MISFET) was demonstrated by Ueno [32] using an undoped SrTiO$_3$ substrate as a conduction channel and Al$_2$O$_3$ as a gate insulator. Another FET device using SrTiO$_3$ as a channel and CaHfO$_3$ as an insulator was reported by Shibuya with amorphous [37] and epitaxial [36] interfaces.

## Electrical Gating of LaAlO$_3$/SrTiO$_3$ structures

There are two common methods for electrical gating of LaAlO$_3$/SrTiO$_3$ structures. The first uses a top gate to modulate the 2DEG carrier density (Figure 3 (a)). This method of gating was used by Jany *et al* to create Schottky-like devices [38]. As shown in Figure 4, the interface of a 3-4 uc-LaAlO$_3$/SrTiO$_3$ heterostructure is directly contacted with a buried contact; an adjacent area is top-gated through the LaAlO$_3$. When forward biased (Figure 4 (a)), the interface becomes conducting and a large current flows. Under negative gate bias (Figure 4 (b)), the 2DEG is depleted and the conduction between the gate and interface contact becomes very low. The top electric field effect switches the interface between insulating and conducting states, leading to highly rectifying behavior (Figure 4 (c)).

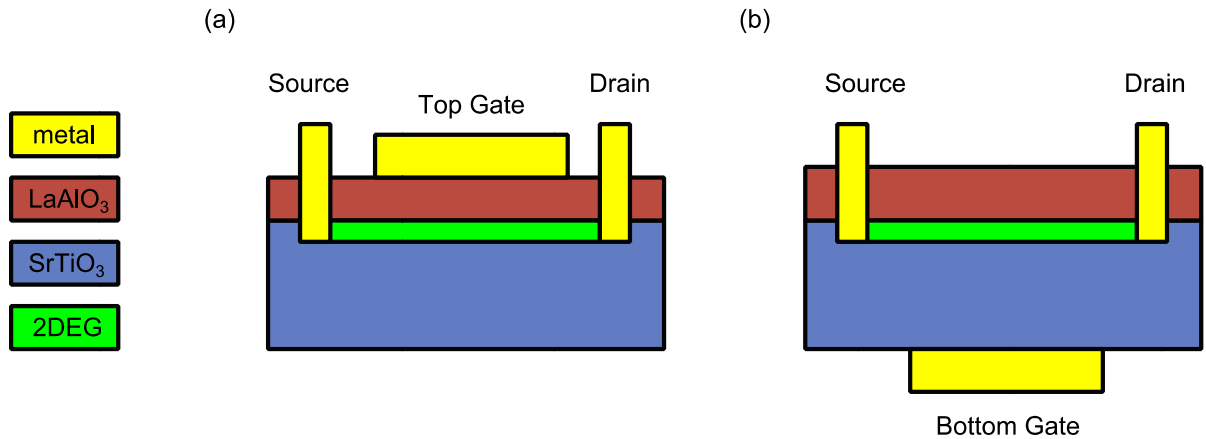

Figure 3. Two methods for electrical gating of LaAlO$_3$/SrTiO$_3$ heterostructures. (a) Top gating, where the electric field is applied across the thin (~1 nm) LaAlO$_3$ layer. (b) Bottom gating, where the electric field is applied across the thick (~1 mm) SrTiO$_3$ layer.

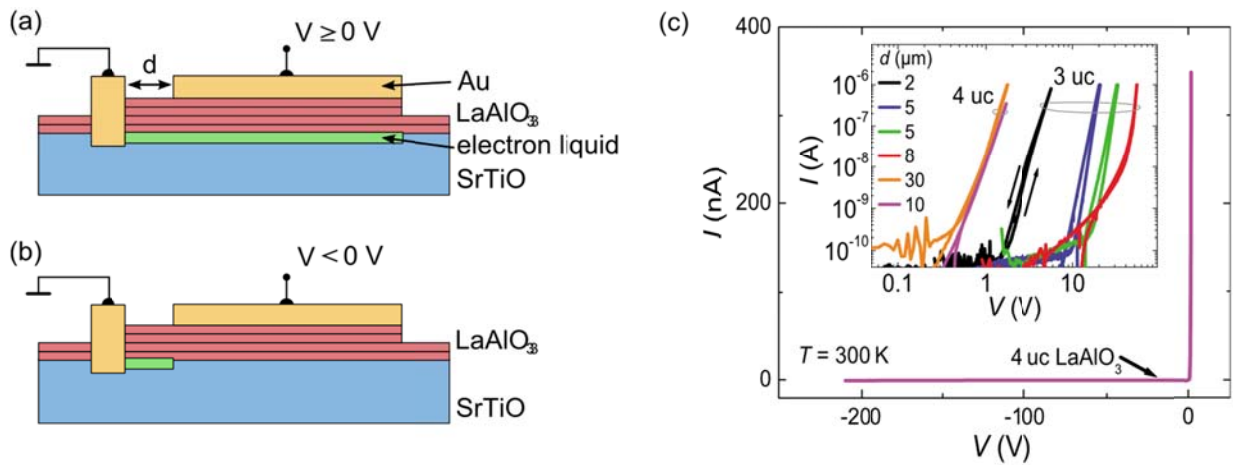

Figure 4. Illustration of a LaAlO$_3$/SrTiO$_3$ field-effect device exhibiting Schottky-diode behavior. (a) Enhancement mode ($V \geq 0$) in which there is large conductance between the interface contact and top gate. (b) Depletion mode ($V < 0$) in which the interface conduction has been extinguished. (c) I-V curve. (Adapted from Ref. [38].)

A second approach to electrical gating is to apply a voltage to the bottom of the relatively thick (~1 mm) SrTiO$_3$ substrate (Figure 3 (b)). Although the effect of a back gate is diminished by the spatial separation from the 2DEG layer, it is simultaneously enhanced by the larger dielectric constant of SrTiO$_3$ (~300) compared to LaAlO$_3$ (~25). The transport properties of the LaAlO$_3$/SrTiO$_3$ 2DEG can be modified by applying large voltages (±100 V) to the bottom of the SrTiO$_3$ substrate. At room temperature, this type of gating can lead to hysteretic switching of 3uc-LaAlO$_3$/SrTiO$_3$ structures [13]. The discovery of interfacial superconductivity at the LaAlO$_3$/SrTiO$_3$ interface was subsequently found to be electric-field tunable using the same type of back gating [39].

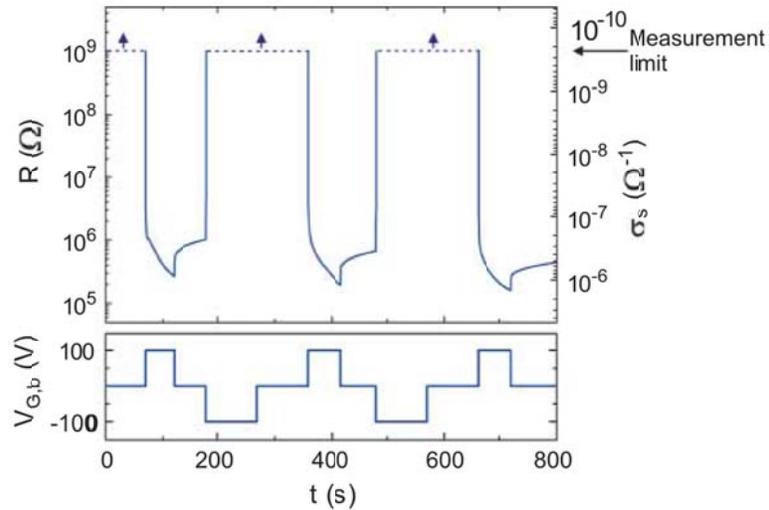

Figure 5. Hysteretic switching of the resistance of 3 uc-LaAlO$_3$/SrTiO$_3$ using a back gate. Positive gate biases produce a conducting interface, even after the bias is restored to zero. Similarly, negative gate biases produce an insulating interface. (Adapted from Ref. [13].)

## LaAlO$_3$/SrTiO$_3$ -based field-effect devices

One of the earliest attempts to laterally define conducting microstructures at the LaAlO$_3$/SrTiO$_3$ interface focused on lithographically modifying the growth process in order to spatially modulate the number of *crystalline* LaAlO$_3$ layers [40]. In the first step, 2 uc LaAlO$_3$ are grown epitaxially on SrTiO$_3$ (Figure 6 (a)). A lithographic mask is then used to define areas on which to grow *amorphous* LaAlO$_3$ (Figure 6 (b)). Following liftoff, an additional 3 uc of *crystalline* LaAlO$_3$ are grown (Figure 6 (c)). The result is conducting interfaces beneath the regions with 5 uc LaAlO$_3$/SrTiO$_3$ surrounded by insulating regions 2 uc LaAlO$_3$/SrTiO$_3$, as shown in Figure 6 (d). Using a combination of optical and electron-beam lithography, conducting features as small as 200 nm were demonstrated.

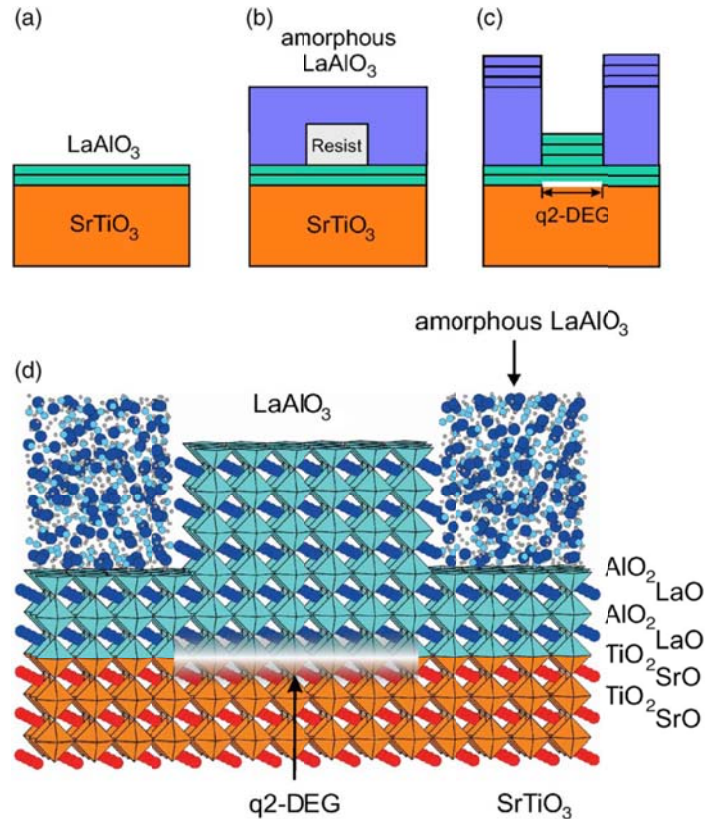

Figure 6. Lithographic method for producing conducting microstructures at the LaAlO$_3$/SrTiO$_3$ interface. (a) Two unit cells of LaAlO$_3$ are grown on TiO$_2$-terminated SrTiO$_3$. (b) Selected areas are masked off by photoresist, and amorphous LaAlO$_3$ is grown. (c) After liftoff, LaAlO$_3$ grows epitaxial on areas that were masked off, but not on the amorphous regions. (d) Final structure, illustrating patterning of the 2DEG. (Adapted from Ref. [40].)

# Reconfigurable nanoscale devices

A powerful method for creating nanoscale devices at the LaAlO$_3$/SrTiO$_3$ interface involves metastable charging of the top LaAlO$_3$ surface with a conducting AFM probe. By locally and reversibly controlling a metal-insulator transition, the creation of both isolated and continuous conducting features has been demonstrated with length scales smaller than 2 nm. These structures can be erased and rewritten numerous times. As a result of the enormous flexibility in controlling electronic properties at near-atomic dimensions, a variety of nanoscale devices can be realized. Here we describe the nanoscale writing technique and illustrate some of these prototype devices and their properties.

### Nanoscale writing and erasing

To create conducting nanostructures, a conducting AFM tip is placed in contact with the top LaAlO$_3$ surface and biased at $V_{tip}$ with respect to the interface, which is held at electrical ground (Figure 7). Positive tip voltages locally produce a metallic interface, while negative tip voltages locally restore the insulating state. During the writing and erasing process, the conductance is monitored between buried Au electrodes that directly contact the interface. Figure 8 (a) illustrates the result of writing with $V_{tip} = +3$ V. A sudden increase in conductance is observed when a conducting path is obtained between the two monitored electrodes (y = 0 nm).

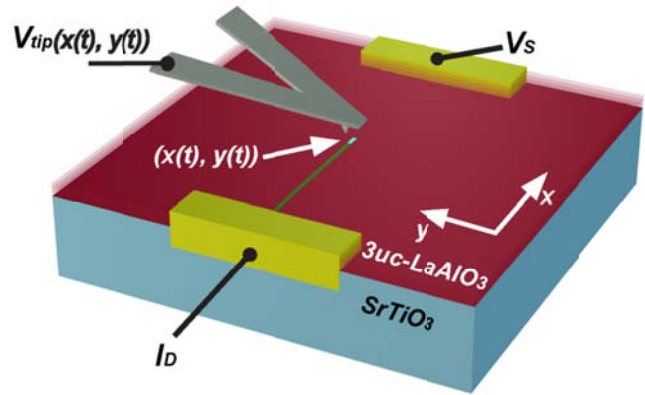

Figure 7. Schematic illustrating of the nanowriting process at the LaAlO$_3$/SrTiO$_3$ interface. Au electrodes (shown in yellow) are electrically contacted to the LaAlO$_3$/SrTiO$_3$ interface. The AFM tip with an applied voltage is scanned once between the two electrodes with a voltage applied $V_{tip}(x(t),y(t))$. Positive voltages locally switch the interface to a conducting state, while negative voltages locally restore the insulating state. Here, a conducting nanowire (shown in green) is being written. The conductance between the two electrodes is monitored by applying a small voltage bias on one of the two gold electrodes ($V_S$) and reading the current at the second electrode ($I_D$). (Adapted from Ref. [41].)

To provide a measure of the transverse dimension of the conducting wire, and to demonstrate that the writing process is reversible, the wire can be "cut" with a reverse voltage $V_{tip} = -3$ V (Figure 8 (b)). A sharp reduction in current is observed, comparable in abruptness to the one found for the writing process. Assuming that the erasure process has a resolution comparable to the writing process, one can infer the nanowire width from the deconvolved differential profile full width at half maximum (FWHM).

The writing and erasing process can be repeated hundreds of times without noticeable degradation of the nanowire. Figure 8 (c) illustrates the repeated erasing and writing of a nanowire using 100 ms voltage pulses. The inset shows that a conducting and insulating state is consistently achieved.

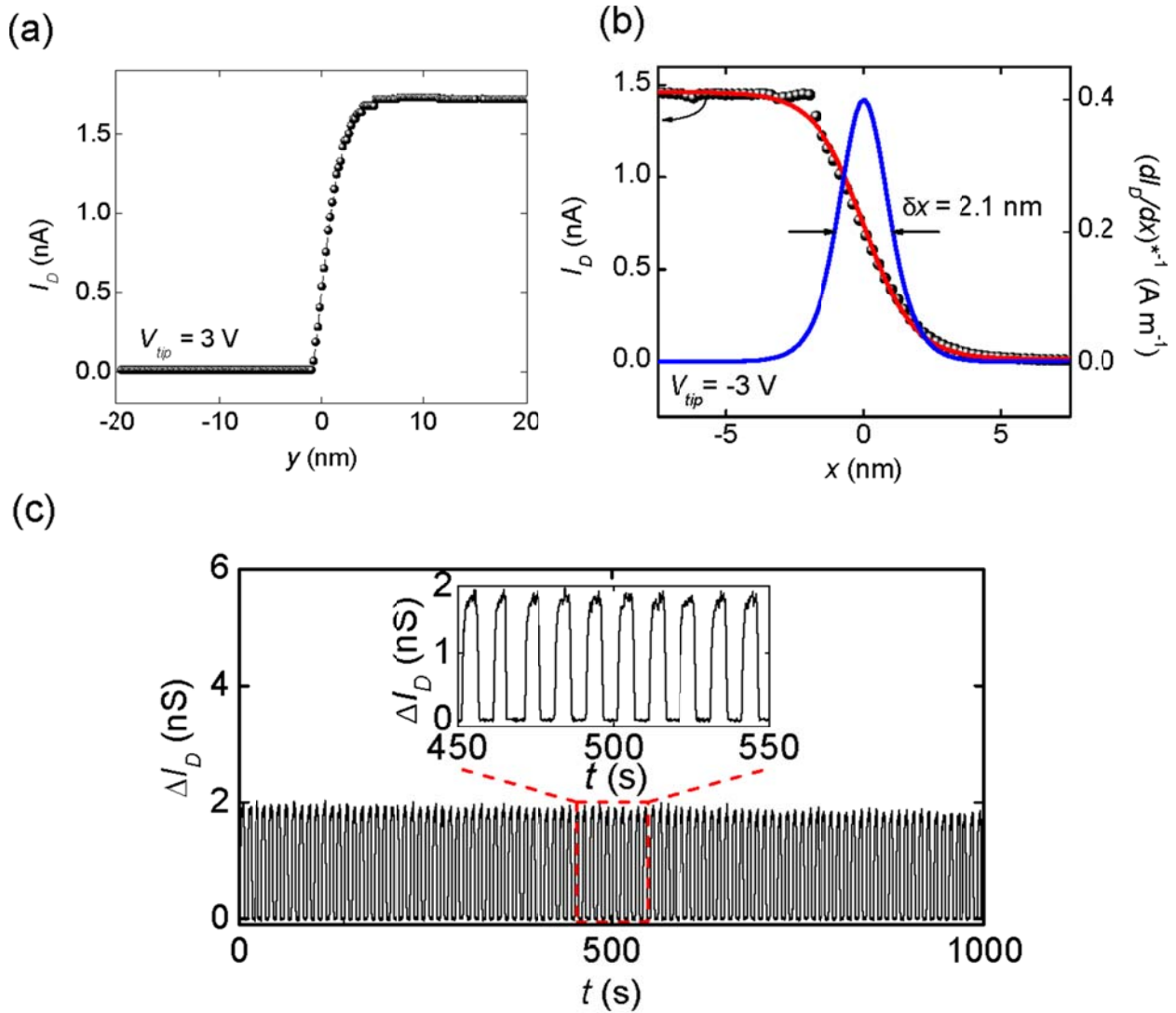

Figure 8. Writing and erasing nanostructures at the LaAlO$_3$/SrTiO$_3$ interface (a) Conductance between the two electrodes measured with a lock-in amplifier as a function of the tip position while writing a conducting wire with $V_{tip}$ = +3 V. A steep increase in conductance shows when the tip reaches the second electrode $\delta x$ = 3.3 nm. (b) Conductance drop when the wire is cut with $V_{tip}$ = -3 V. (c) Repeated cutting and restoring the conductance of a 12 nm-wide nanowire using $V_{tip}$ = ±10 V. (Adapted from Ref. [42] and [43].)

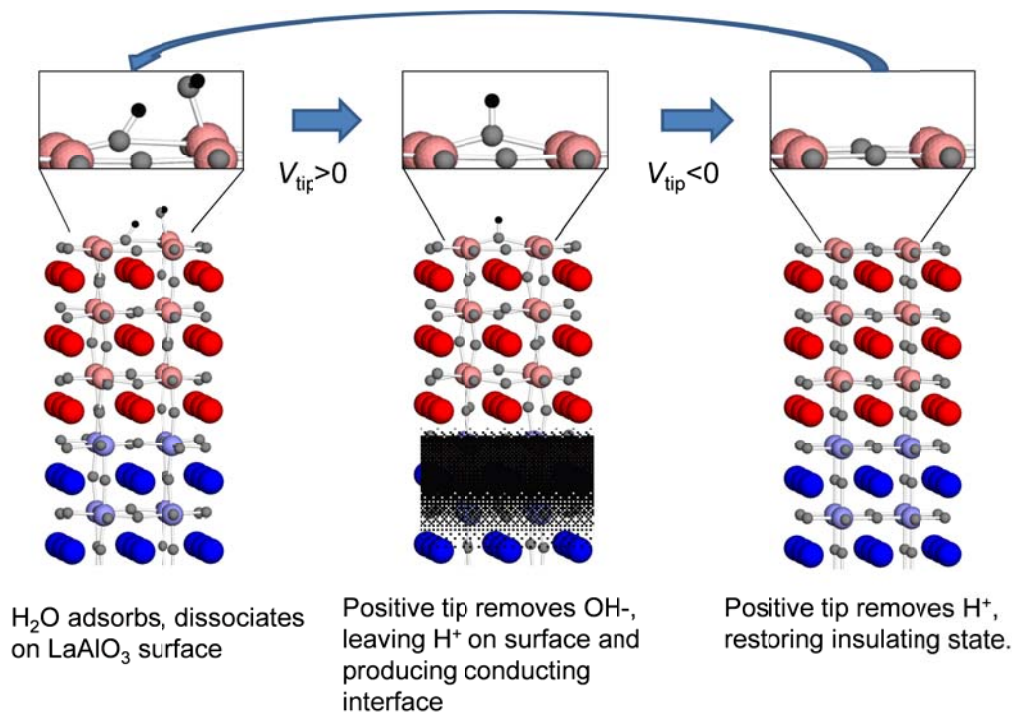

Figure 9. "Water cycle" mechanism for writing and erasing structures at the $LaAlO_3/SrTiO_3$ interface. (Courtesy, C. S. Hellberg).

## "Water cycle"

One possible explanation for the switching of the 2DEG interface between the insulating and conducting states is the selective removal of $OH^-$ or $H^+$ in the water layer naturally adsorbed at $LaAlO_3$ top surface. We can assume that the top surface has naturally water adsorbed on the top $AlO_2$ layer in atmosphere. The water is dissociated in $OH^-$ and $H^+$ as illustrated in Figure 9. The large binding energy makes the first layer stable even in ultra-high vacuum. When the positively charged AFM tip scans the top surface, the $OH^-$ will be removed and leave behind a positive $H^+$. Scanning with a negative voltage will remove the remaining $H^+$ and thus restore the neutral state of the surface. The neutral surface will easily adsorb new water molecules.

## $LaAlO_3/SrTiO_3$ as floating-gate transistor network

The properties of $LaAlO_3/SrTiO_3$ heterostructures are similar to enhancement-mode MOSFET devices used to form logic gates in microprocessors and other digital electronics. Figure 10 (a, b) illustrates the "off" and "on" states of an n-type, enhancement-mode MOSFET. In the off state, electrons cannot form a conducting path from source to drain because the region below the gate is p-type. However, application of a positive bias to the gate can create a channel at the oxide-silicon interface (Figure 10 (b)), thus closing the switch. The switch is volatile, in that when the externally applied voltage is turned off, the switch returns to its open state. A 3 uc-$LaAlO_3/SrTiO_3$ heterostructure can behave in a similar fashion. Figure 10 (c, d) illustrates the

"off" and "on" states of a LaAlO$_3$/SrTiO$_3$ heterostructure. The main difference is that the top gate is "floating" and can store charge metastably. This charge can be added and removed with extremely high spatial resolution using a scanning probe microscope. The 3 uc-LaAlO$_3$/SrTiO$_3$ system can thus be regarded as a two-dimensional network of non-volatile field-effect transistors. As will be seen below, the patterns of charge applied to the top LaAlO$_3$ surface can be used to create conducting lines, isolated islands, transistors, and other devices.

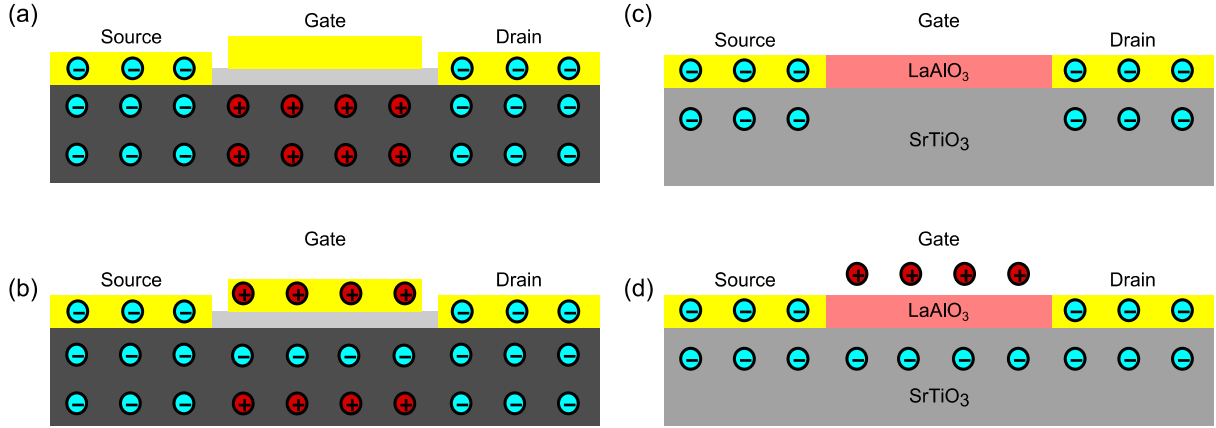

Figure 10. Enhancement-mode MOSFET in (a) "off" and (b) "on" states and LaAlO$_3$/SrTiO$_3$ device in (c) "off" and (d) "on" states.

## Quasi-0D Structures

It is also possible to write isolated conducting islands or "dots" by applying voltage pulses with fixed amplitude but varying duration at a fixed tip position. A monotonically increasing island size with pulse duration is observed in writing experiments that create a chain of dots spaced a fixed distance apart. The critical spacing for transitioning between isolated dots and a continuous line can vary from 1 nm~ to > 35 nm depending on the pulse duration [43].

## Designer Potential Barriers

The high degree of control over the energy landscape within the LaAlO$_3$/SrTiO$_3$ 2DEG allows for the development of a variety of nonlinear devices such as nanoscale junctions [43]. The shape of the barrier can determine whether the transport is reciprocal ($I(V) = -I(-V)$) or rectifying.

The controlled creation of rectifying structures is further described below. Non-reciprocal nanostructures can be created using a slightly different c-AFM manipulation. In this approach, spatial variations in the conduction-band profile are created by a precise sequence of erasure steps. In a first experiment, a conducting nanowire is created using $V_{tip} = +10$ V. The initial $I$-$V$ curve (Figure 11 (a), green curve) is highly linear and reciprocal. This nanowire is then cut by scanning the AFM tip across the nanowire at a speed $v_y = 100$ nm/s using $V_{tip} = -2$ mV at a fixed location (x = 20 nm) along the length of the nanowire. This erasure process increases the conduction-band minimum $E_c$ (x) locally by an amount that scales monotonically with the number of passes $N_{cut}$ (Figure 11 (a), inset); the resulting nanostructure exhibits a crossover from conducting to activated to tunneling behavior [41]. Here we focus on the symmetry of the full I-V curve. As $N_{cut}$ increases, the transport becomes increasingly nonlinear; however, the I-V curve remains highly reciprocal. The canvas is subsequently erased and a uniform conducting nanowire is written in a similar fashion as before ($V_{tip}$ = +10 V, $v_x$ = 400 nm/s). A similar erasure sequence is performed; however, instead of cutting the nanowire at a single x coordinate, a sequence of cuts is performed at nine adjacent x coordinates along the nanowire (separated by $D_x$ = 5 nm). The number of cuts at each location along the nanowire $N_{cut}$ (x) increases monotonically with x, resulting in a conduction band profile $E_c$ (x) that is asymmetric by design (Figure 11 (b), inset). The resulting I-V curve for the nanostructure evolves from being highly linear and reciprocal before writing (Figure 11 (b), (green curve)) to highly nonlinear and non-reciprocal (Figure 11 (b), red curve).

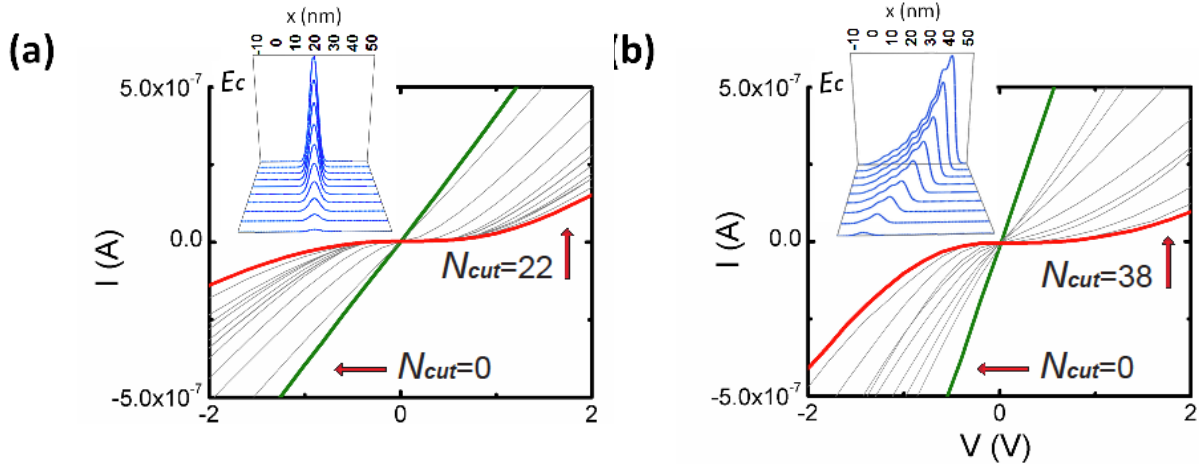

Figure 11. (a) I-V plots for a nanowire cut at the same location multiple times with an AFM tip bias $V_{tip}$ = -2 mV. The green curve indicates the I-V curve before the first cut. Intermediate I-V curves are shown after every alternate cut. As the wire is cut, the potential barrier increases (inset) and the zero-bias conductance decreases; however, the overall I-V curve remains highly reciprocal. (b) I-V plots for a nanowire subject to a sequence of cuts $N_{cut}$ (x) at nine locations spaced 5 nm apart along the nanowire (see Table I). The green curve indicates I-V curve before the first cut. The asymmetry in $N_{cut}$ (x) results in a non-reciprocal I-V curve. (Adapted from Ref. [41].)

Nanoscale control over asymmetric potential profiles at the interface between LaAlO$_3$ and SrTiO$_3$ can have many potential applications in nanoelectronics and spintronics. Working as straightforward diodes, these junctions can be used to create half-wave and full-wave rectifiers for AC-DC conversion or for RF detection and conversion to DC. By cascading two or more such junctions, with a third gate for tuning the density in the intermediate regime could form the basis for low-leakage transistor devices. The ability of controlling the potential along a nanowire could be used to create wires with built-in polarizations similar to those created in heterostructures that lack inversion symmetry [44].

## SketchFET

Here we describe the creation of a sketch-defined transistor or "SketchFET". A transistor is a three-terminal device and one begins with a uniformly non-conductive region that spans three electrodes (Figure 12 (a)). A "T-junction" is written which will form the source (S), gate (G), and drain (D) leads of the SketchFET ((Figure 12 (b)). The central junction is then erased over a ~1μm area (Figure 12 (c)), and a smaller junction is created using a narrower line width (Figure 12 (d)). A close-up of this region is shown in Figure 12 (e). Finally, the horizontal nanowire is cut and the gate electrode is moved further away (~25 nm).

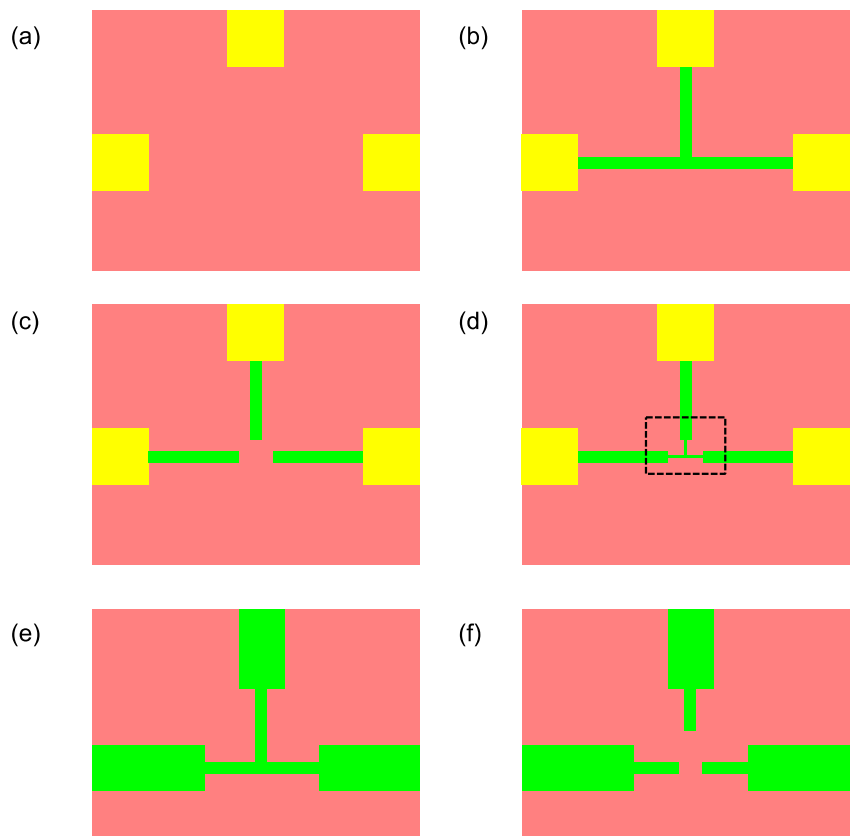

Figure 12. Schematic illustration of step-by-step creation of a SketchFET.

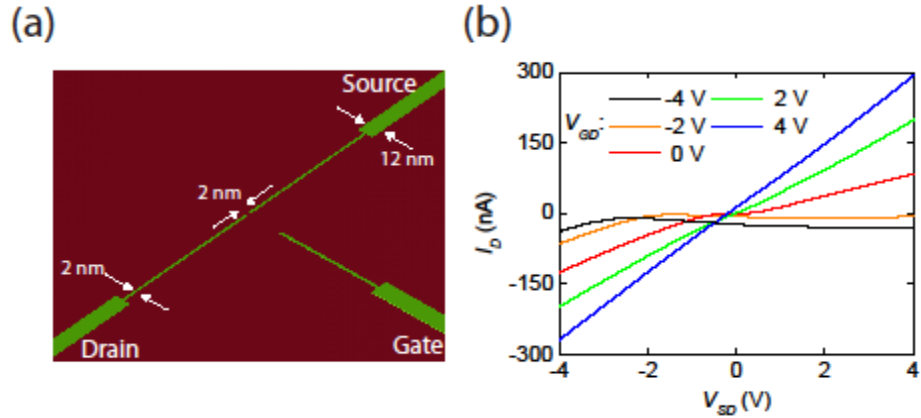

Figure 13. SketchFET device characteristics. (a) Schematic diagram of SketchFET structure. (b) *I-V* characteristic between source and drain for different gate biases $V_{GD}$. Adapted from Ref. [45].

Transport measurements of a SketchFET can be performed by monitoring the drain current $I_D$ as a function of the source and gate voltages ($V_{SD}$ and $V_{GD}$, respectively). Both $V_{SD}$ and $V_{GD}$ are referenced to the drain, which is held at virtual ground. At zero gate bias, the *I-V* characteristic between source and drain is highly nonlinear and non-conducting at small $|V_{SD}|$ (Figure 13 (b)). A positive gate bias $V_{GD}>0$ lowers the potential barrier for electrons in the source and gate leads. With $V_{GD}$ large enough (>=5V in this specific device), the barrier eventually disappears. In this regime, ohmic behavior between source and drain is observed. This field effect in this case is non-hysteretic, in contrast to field effects induced by the AFM writing procedure. When a sufficiently large gate bias is applied, a small gate leakage current $I_{GD}$ can contribute to the total drain current $I_D$. The amount of leakage current can be adjusted by placing the gate electrode at different distances away from the source-drain junction. Conductance changes of up to four orders of magnitude have been observed in SketchFET devices, which make them interesting components for logic devices. We note that complementary logic families directly analogous to CMOS are not possible since only electron gases have been demonstrated thus far. A significantly different logic device must be constructed that can simultaneously switch its logic state and consume minimal power while holding either of two states.

**Nanoscale photodetectors**

In addition to the variety of electronic devices described above, one can also use AFM patterning of the LaAlO$_3$/SrTiO$_3$ interface to create rewritable, nanoscale photodetectors. Nanophotonic devices seek to generate, guide, and/or detect light using structures whose nanoscale dimensions are closely tied to their functionality [46, 47]. Nanoscale photodetectors created at the interface of LaAlO$_3$/SrTiO$_3$ possess an electric-field tunable spectral response spanning the visible-to-near-infrared regime. Following illumination with up to kW/cm$^2$ of optical intensity they are still able to be erased and rewritten. An analysis of noise equivalent target (NET)) shows a minimum NET of 11 mW/cm$^2$/$\sqrt{Hz}$ (at T = 80 K and $\lambda$ = 735 nm).

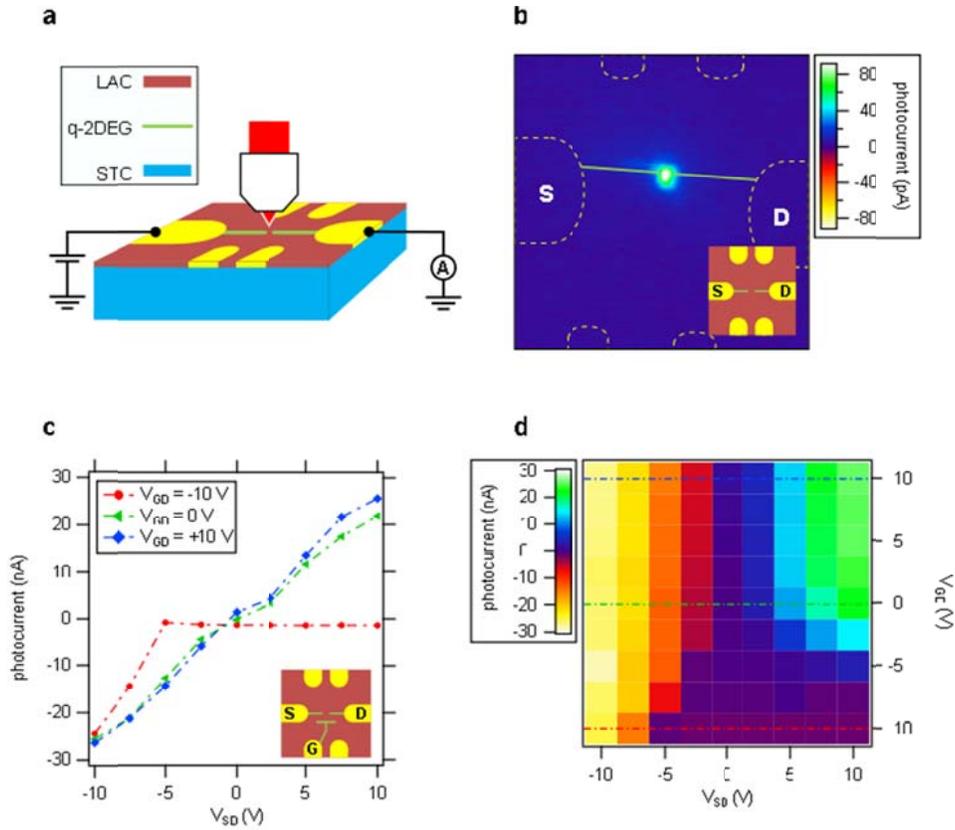

Figure 14 (a) Diagram of photocurrent measurement. (b) Scanning photocurrent microscopy (SPCM) image of two-terminal device shown in inset. $I \sim 30$ kW/cm$^2$ (NA = 0.73), $V_{SD} = 0.1$ V, $T = 300$ K. (c) Photocurrent vs. $V_{SD}$ of the three-terminal device shown in the inset. $I \sim 20$ W/cm$^2$ (NA = 0.13), $T = 80$K. (d) Intensity map of photocurrent of three-terminal device as a function of $V_{SD}$ and $V_{GD}$.

Optical properties of nanostructures are spatially mapped using scanning photocurrent microscopy (SPCM) [48, 49] (Figure 14 (a)). The intensity of a laser source is modulated by an optical chopper at frequency $f_R$ and the resulting photocurrent is measured with a lock-in amplifier at $f_R$. When the light overlaps with the device a sharp increase in the photocurrent is observed. An SPCM image shows spatially-localized photocurrent detected only in the region of the junction (Figure 14 (b). In a three-terminal geometry (née SketchFET) the gate-drain bias ($V_{GD}$) can be used to modify the source-drain conductance, enabling conduction between source and drain for positive $V_{GD}$ and inhibiting it for negative $V_{GD}$ (Figure 14 (c)).

The spectral response is sensitive to both $V_{SD}$ and $V_{GD}$. At positive $V_{SD}$ the photodetector response red-shifts as the gate bias is increased. A similar Stark shift is observed when sweeping the source bias. This evidence of a Stark effect, along with finite element analysis showing that the electric field is predominantly confined to the gap region, indicates that the photo-induced absorption is highly localized.

## Integration of LaAlO$_3$/SrTiO$_3$ with Silicon

Because of its ubiquitous use in modern microelectronics, silicon is the most desirable platform on which to base multifunctional oxide-based nanoelectronic devices. As a result, one of the major challenges of oxide-based electronic device research is integrating with Si-based electronic circuits and subsequently scaling to a commercially relevant available large wafer process compatible with present Si-process. Using a combination of oxide growth techniques, the integration of oxide heterostructures with silicon can be achieved. Oxide molecular beam epitaxy (MBE) growth of virtual $SrTiO_3$ substrate layers can be achieved on Si substrates, followed by pulse laser deposition (PLD) or laser-MBE growth of oxide heterostructures [50].

Figure 15 shows cross-sectional TEM images of oxide heterostructures consisting of 5 uc of $LaAlO_3$ grown on a 100 nm-thick $SrTiO_3$ template grown on Si substrate. The 100 nmthick epitaxial (001) $SrTiO_3$ layers were deposited on (001) Si wafers in an 8 inch-diameter molecular beam epitaxy (MBE) machine [50]. The termination of the epitaxial $SrTiO_3$ thin films was controlled by halting the film growth at $SrO^-$, $TiO_2^-$, or undefined-termination. During the STO deposition, if surface termination is not controlled, it may have both $SrO^-$ and $TiO_2^-$ terminations. To improve the crystalline quality and surface morphology, all $SrTiO_3$-on-Si templates were annealed at 900 °C for 2 hrs in $O_2$ atmosphere [51].

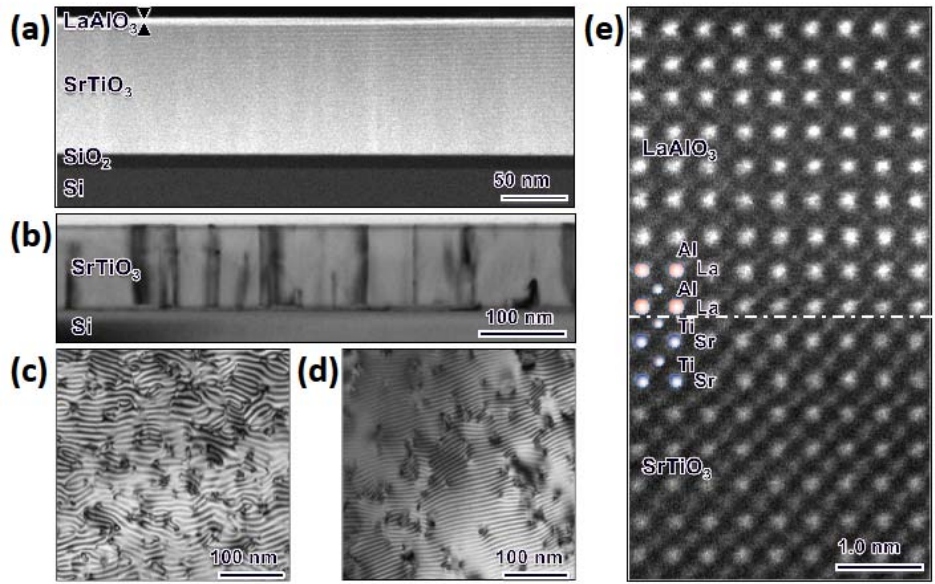

Figure 15 TEM investigation of $LaAlO_3$/$SrTiO_3$ heterointerface on Si. (s) Cross-sectional HAADF image of 5 nm-thick $LaAlO_3$/annealed $TiO_2^-$ $SrTiO_3$ heterostructure grown on Si. (b) Cross-sectional TEM image of the same sample showing the existence of threading dislocations in the $SrTiO_3$ layer. (c,d) Planar view TEM images of (c) as-grown and (d) annealed $SrTiO_3$ films showing moiré patterns. (e) High-resolution HAADF image showing an atomically sharp interface between LaAlO3 film and annealed $TiO_2^-$ $SrTiO_3$ on Si. (Adapted from [51].)

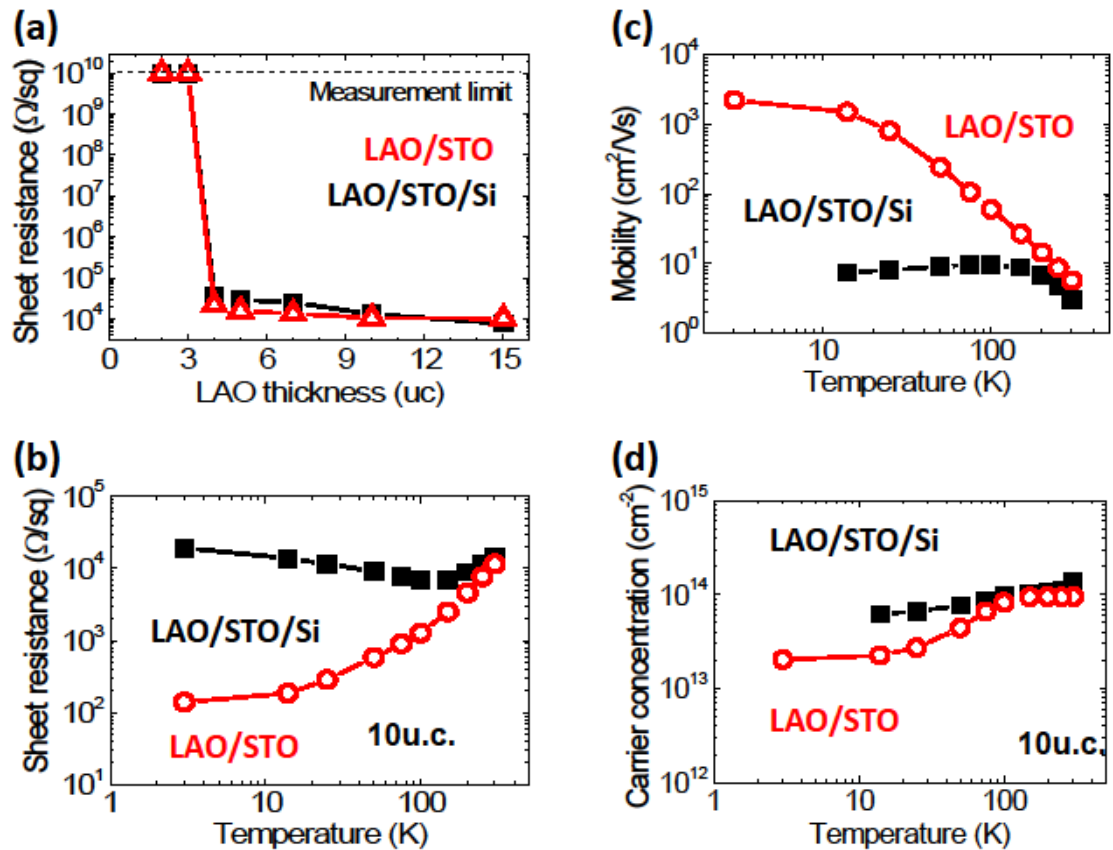

Figure 16. Electrical transport properties of LaAlO$_3$/SrTiO$_3$ heterointerface on Si. Temperature dependence of (a) sheet resistance, (b) carrier concentration, and (c) mobility of the heterointerface between 10 uc LaAlO$_3$ and TiO$_2$-SrTiO$_3$ on Si. The SrTiO$_3$ layer on Si substrate without LaAlO$_3$ layer showed highly insulating behavior, indicating that the measured transport properties of LaAlO$_3$/SrTiO$_3$ on Si originated only from the 2DEG. For comparison, the electrical transport properties of 10 uc LaAlO$_3$ on SrTiO$_3$ single crystal were also measured. (d) LaAlO$_3$ thickness dependence of sheet resistance of unpatterned LaAlO$_3$/ SrTiO$_3$ heterointerface on Si. (Adapted from [51]).

The temperature dependence of the sheet resistance, carrier concentration, and mobility of an unpatterned heterointerface between 10 uc LaAlO$_3$ and annealed TiO$_2$-SrTiO$_3$ on Si are measured by the van der Pauw method (Figure 16). The room-temperature properties are comparable to those for heterointerfaces between LaAlO$_3$ and TiO$_2$terminated SrTiO$_3$ [5, 52]. Although the low-temperature mobility is lower than LaAlO$_3$/SrTiO$_3$ single crystal, presumably due to the sensitivity to the defect structure of the heterointerface, we believe that the 2DEG at the LaAlO$_3$/SrTiO$_3$ heterointerface on Si is useful for room-temperature nanoelectronic devices. A comparison between the room-temperature thickness-dependent electrical properties of unpatterned LaAlO$_3$/SrTiO$_3$ on Si and LaAlO$_3$ on SrTiO$_3$ bulk single crystal grown under identical conditions shows the same 4 uc critical thickness. The electrical properties of the SrTiO$_3$ layer on Si shows insulating behavior relative to the interfacial 2DEG for LaAlO$_3$ thicknesses above the critical thickness. The bare Si substrate was determined to be *p*-type, with a room-temperature mobility $\mu \sim 350$ cm$^2$/Vs and carrier concentration $n \sim 0.8 \times 10^{15}$ cm$^{-3}$, significantly different from the electrical properties of the interfacial 2DEG for LaAlO$_3$/SrTiO$_3$ on Si.

Nanostructure writing experiments show that reversible patterning of the LaAlO$_3$/SrTiO$_3$ interface can be achieved in essentially the same way as for LaAlO$_3$/SrTiO$_3$ interfaces formed on bulk SrTiO$_3$ substrates. The ability to sustain 2DEG behavior depends on the interface preparation. 2DEG behavior has not been found for the heterointerfaces between 3 uc LaAlO$_3$ and SrO- or undefined-terminated SrTiO$_3$ templates on Si. These results are consistent with previous reports that a well-defined TiO$_2$-terminated SrTiO$_3$ surface is critical for the formation of the 2DEG. Even the heterointerface between LaAlO$_3$ on as-grown TiO$_2$- SrTiO$_3$/Si did not exhibit 2DEG behavior [51] and it is expected that his result is due to the defective surface of the as-grown SrTiO$_3$ template on Si substrate. In contrast, the annealed SrTiO$_3$ template has relatively lower dislocation density through the annihilation of dislocations with contrary Burgers vectors and the dissociation of two whole dislocations into partial dislocations during high-temperature post-annealing [53]. Further optimization of the SrTiO$_3$/Si substrate quality is needed for useful device applications.

# Future Prospects

It is still far too premature to make definitive proclamations about the applicability of this novel interface for technology, but there are many interesting features that are certainly worth exploring. There are well-defined possibilities regarding ultra-high-density transistor and memory elements, though stability and scaling issues are just as clearly defined. In keeping with Kroemer's lemma, it is important to keep an open mind to entirely new classes of devices that are enabled by the unique combination of properties present in the LaAlO$_3$/SrTiO$_3$ system. Several possible directions are discussed below.

### Room-Temperature Devices

Semiconductors are the workhorse of modern electronics industry. To obtain ever higher device integration densities, the size of metal-oxide-semiconductor field-effect transistors (MOSFET) have successfully followed Moore's Law for over four decades. Now, due to the intrinsic limitations, the scaling of MOSFET devices is truly reaching fundamental limits [54-56]. In addition to being a possible avenue for scaling beyond Moore's law, oxide-based 2DEG devices may also provide opportunities in terms of adding new functionality.

### Information Processing

SketchFET structures are interesting candidates for post-CMOS transistors. The quantum tunneling-dominated field emission operates in a distinct manner compared with traditional MOSFET devices. Due to the fact that the gate, source, drain, and junction areas are essentially made of the same material, SketchFETs are less affected by the "short channel" effect which is a major limit of the MOSFET scaling [57].

AFM lithography eliminates defects that can be introduced during conventional chemical lithography steps and also obtains the separation of carriers from dopants through modulation doping processes. These achievements, coupled with future improvements of material quality and device layout, have the potential to produce very high carrier mobility and lead to SketchFETs with fast switching speeds. Further enhancements in the room-temperature mobility of devices may be possible through advances in materials growth techniques.

Scaling to large numbers of SketchFET transistors will require either a scalable array of AFM probes (e.g., IBM's Millipede project [58]) or other techniques for creating the desired $LaAlO_3$ surface state. Nanoimprint lithography [59] may prove useful for the latter approach. Still, it will be important to integrate oxide materials with conventional silicon-based electronics.

Using an AFM to lithographically pattern oxide interfaces allows the fabrication and integration of devices with different functionalities on a single chip without the need to incorporate many different materials. Junctions between nanowires can be formed with various potential profiles and can be cascaded to produce, for example, nanodiodes with different polarities and turn-on voltages. In the presence of mid-bandgap states such as oxygen vacancies, junctions can be placed and electrically gated to form wide bandwidth nanoscale photodetectors, allowing on-chip optical signal coupling. The nanoscale footprint of such photodetectors could be especially useful in near field optics. In addition, nanoscale control over inversion symmetry breaking in junction profiles could in principle be used to produce nonlinear optical frequency conversion (i.e., second-harmonic generation or difference frequency mixing), thus providing a means for the generation of nanoscale sources of light or THz radiation.

**Spintronics**

Taking advantage of the large Rashba coupling in these 2DEG structures may enable new classes of spintronic devices. Intrinsic magnetic effects have already been reported at temperatures as high as 35 K [52, 60, 61], and large spin-orbit splittings have been measured through magnetotransport [44, 62, 63]. The resulting effective magnetic fields could allow control over spin precession along two orthogonal axes [64] and thus exert full three-dimensional control over electron spin [65] in a nanowire. Another approach to spintronics could be in the development of a Datta-Das spin transistor [66], which would require spin injection and filtering components. The development of spin-based oxide nanostructures could lead to hybrid magnetoelectric components with low power requirements.

## Quantum Hall regime

For almost three decades, the integer and fractional quantum Hall effects has been observed almost exclusively in high-mobility silicon or III-V heterostructures. Recently, Tsukazaki *et al*. have observed both the integer [3] and fractional quantum Hall effect [67] in high mobility ZnO/Mg$_x$Zn$_{1-x}$O heterostructures. In the case of a planar, well-oxidized LaAlO$_3$/SrTiO$_3$ 2DEG, the observation of quantum Hall phenomena seems *a priori* precluded by its low carrier mobility and high carrier density. However, the carrier density can be much lower in AFM-written nanowires than in planar conductive LaAlO$_3$/SrTiO$_3$ interfaces. Secondly, quasi-one-dimensional confinement from the nanowire is expected to suppress elastic backscattering and increase carrier mobility [68]. A third effect arises from the peculiar dielectric properties of SrTiO$_3$ at low temperature: the large local electric field inside the nanowire formed by writing process can lead to a dielectric constant in the nanowire that is significantly less than the surrounding region. For this geometry, Jena *et al* predict an order-of-magnitude mobility enhancement in two dimensions and additional enhancements for one-dimensional geometries [69]. Nanowires created via conducting-AFM with controllable carrier concentration and mobilities may be suitable for quasiparticle interference experiments in experimentally realizable topological quantum computing geometries [70, 71].

## Superconducting Devices

SrTiO$_3$ was one of the first semiconductors discovered to exhibit superconducting behavior at low temperature [72, 73]. Recently, superconductivity in SrTiO$_3$ has experienced a renaissance due to the discovery of 2D superconductivity at the LaAlO$_3$/SrTiO$_3$ interface [74]. The measured transition temperature $T_C$ is approximately 200-400 mK (which is comparable to bulk superconducting SrTiO$_3$) and is LaAlO$_3$-thickness dependent [74]. Subsequent investigations have shown that the application of an electric field can tune the carrier density and induce a superconductor-metal quantum phase transition [39].

In addition to constructing superconducting wires and Josephson couplings, new types of interactions may be possible that take advantage of both the dimensionality of those structures and other intrinsic properties like the strong spin-orbit coupling in these systems [44].

## Solid state Hubbard simulators

One long-range goal might be the creation of a platform for quantum simulation—a *solid-state* Hubbard toolbox—using the LaAlO$_3$/SrTiO$_3$ interface whose metal-insulator transition can be controlled on scales approaching the underlying unit cell (Figure 17). The ground states of strongly correlated systems have generally eluded exact analytical solution. For example, the phase diagram of the 2D Hubbard model away from half filling, which has been proposed to explain the mechanism for high-temperature superconductivity, is still poorly understood.

Numerical methods remain limited in their ability to explore the full relevant phase space. Some novel approaches to this problem involve the use of physical simulations using atomic gases trapped in optical lattices [75-79]. Using a so-called "cold atom Hubbard toolbox" [80, 81], it may be possible to explore parameter regimes or measurement classes not accessible in conventional solid-state systems. Pioneering experiments such as the observation of a reversible quantum phase transition between a Bose-Einstein condensate to a Mott insulator [82] lead credence to the idea that quantum systems can simulate one another. In the language of quantum information, optical lattices can be regarded as approaching the universal quantum simulator first conceived by Feynman [83, 84].

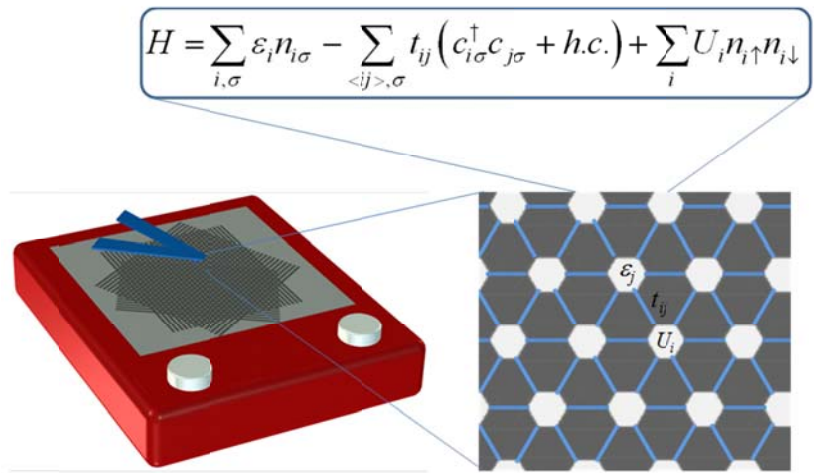

$$H = \sum_{i,\sigma} \varepsilon_i n_{i\sigma} - \sum_{<ij>,\sigma} t_{ij}\left(c_{i\sigma}^{\dagger} c_{j\sigma} + h.c.\right) + \sum_i U_i n_{i\uparrow} n_{i\downarrow}$$

Figure 17. Illustration of $LaAlO_3$/ $SrTiO_3$ as a rewritable nanoelectronic interface. Here, a conducting AFM creates a hexagonal array of conducting islands separated by ultrathin tunnel barriers. Using a demonstrated nanoscale electronic confinement with 2 nm resolution, artificial condensed matter systems such as this may be constructed and probed. For these systems, electronic correlations play a critical role in determining their properties (metallic, insulating, and superconducting).